\documentclass[aps,twocolumn,longbibliography]{revtex4-1}

\usepackage[utf8]{inputenc}
\usepackage{amsmath,amssymb,latexsym,epsfig,graphics,epsf}
\usepackage{color} 
\usepackage[hidelinks]{hyperref}
\usepackage{dcolumn}
\renewcommand{\i}{\operatorname{i}} 
\newcommand{\bR}{\mathbf R}
\newcommand{\bRa}{{\mathbf R}_{\rm a}}
\newcommand{\bRb}{{\mathbf R}_{\rm b}}
\usepackage[nopostdot,nogroupskip,nonumberlist]{glossaries}
\newacronym{WCA}{WCA}{Weeks-Chandler-Andersen}
\newacronym{AWC}{AWC}{Andersen-Weeks-Chandler}
\newacronym{BH}{BH}{Barker-Henderson}
\newacronym{PY}{PY}{Percus-Yevic}
\newacronym{HS}{HS}{hard-sphere}
\newacronym{HH}{HH}{half-harmonic}
\newacronym{FCC}{FCC}{face centered cubic}

\begin{document}

\title{Comparing zero-parameter theories for the WCA and harmonic-repulsive melting lines}
\author{Jeppe C. Dyre}
\author{Ulf R. Pedersen}
\email{ulf@urp.dk}
\affiliation{{\it Glass and Time}, IMFUFA, Department of Science and Environment, Roskilde University, P. O. Box 260, DK-4000 Roskilde, Denmark} 
\date{\today}

\begin{abstract}
The melting line of the Weeks-Chandler-Andersen (WCA) system was recently determined accurately and compared to the predictions of four analytical hard-sphere approximations [Attia \textit{et al.}, J. Chem. Phys. \textbf{157}, 034502 (2022)]. Here, we study an alternative zero-parameter prediction based on the isomorph theory, the input of which relate to properties at a single reference state point on the melting line. The two central assumptions made are that the harmonic-repulsive potential approximates the WCA potential and that pair collisions are uncorrelated. The new approach gives excellent predictions at high temperatures, while the hard-sphere-theory based predictions are better at lower temperatures. Supplementing the WCA investigation, the face-centered-crystal to fluid coexistence line is determined for a system of harmonic-repulsive particles and compared to the zero-parameter theories. The results indicate that the excellent isomorph-theory predictions for the WCA potential at higher temperatures may be partly due to a cancellation of errors between the two above-mentioned assumptions.
\end{abstract}

\maketitle

\section{Introduction}
A central result of classical theories of liquids is that fluid states' structure, dynamics, and statistical properties are determined mainly by short-ranged repulsive forces, while long-range attractions play second fiddle \cite{Bernal1964, Widom1967, Andersen1971, Chandler1983, Berthier2009, Pedersen2010, Hansen2013, Dell2015,  Dyre2016, Chattoraj2020,  Nandi2021, Singh2021, Toxvaerd2021}. This simplification is behind the success of classical hard-sphere (HS) theories \cite{Barker1967, Barker1976, Andersen1971, Heyes2006, Attia2021} and, alternatively, the mapping to harsh effective inverse power-law potential \cite{Hoover1971, Stishov1975, Young1977, Hummel2015}. Both approaches predict structure, dynamics, and statistical properties that are effective functions of a single parameter in the two-dimensional thermodynamic phase diagram. For HS theories the single parameter is the effective HS packing fraction. Several criteria have been suggested for determining this quantity for a given state point (and model); this is in general a subtle problem, the solution of which \textit{de facto} determines the configurational adiabats of the system in question \cite{Rosenfeld1977, Rosenfeld1999}. The oldest approximation for determining the effective HS radius dates back to Boltzmann's thesis \cite{Boltzmann1890}, which proposed that the HS diameter is the shortest distance obtained by two particles colliding head-on with average thermal velocity. In the 1960's, Barker \& Henderson, Andersen \& Weeks \& Chandler, and others developed successful theories based on more rigorous thermodynamic arguments. 

Since the onset of the present millennium, high-pressure experiments by several groups have established that liquid dynamics is often a function of a single thermodynamic parameter. This has been motivated by mapping to a harsh inverse power-law pair potential \cite{Roland2005, Gundermann2011}. The isomorph theory generalizes the idea of an effective single-parameter phase diagram if the potential energy function $U(\bR)$ obeys hidden scale invariance \cite{Gnan2009, Schroeder2014}. This is the property that the ordering of configurations according to their potential energy at one density is maintained upon a uniform scaling of all coordinates, expressed in the logical implication
$U(\bRa)<U(\bRb)\Rightarrow U(\lambda\bRa)<U(\lambda\bRb)$ \cite{Schroeder2014}. While the isomorph theory is mathematically more abstract than the classic HS ideas, it has the advantage of not referring to a specific mapping of pair interactions by viewing the energy landscape in a more holistic way.

The above-mentioned approaches lead to zero-parameter predictions of the shape of the solid-fluid coexistence line for a given system. Here, we study the harmonic-repulsive \cite{Zhu2011, Mohanty2014, Levashov2017, Xu2019, MartnMolina2019, Levashov2019, Levashov2020, Santo2021} and the Weeks-Chandler-Andersen (WCA) \cite{Andersen1971, Nasrabad2008, Ahmed2009, BenAmotz2004b, Benjamin2015, Valds2018, Nandi2021, Singh2021, Banerjee2021, Zhou2022} pair potentials, both of which approach the hard-sphere potential at low temperatures. We can use any reference state point for the isomorph theory of melting \cite{Pedersen2016}. An attractive possibility that removes this arbitrariness is to use the zero-temperature limit as reference state point, which as shown below results in a theoretical prediction close to that of HS theories. We derive the isomorph-theory prediction for an arbitrary finite-temperature reference state point and compare it to a prediction based on classic HS ideas. While the HS approach gives the best predictions at extremely low temperatures, the isomorph theory is more accurate at higher temperatures. 

We first introduce the two energy surfaces of interest (Sec.\ \ref{sec:potentials}). The low-temperature solid-fluid coexistence line of the WCA system was determined accurately in Ref.\ \cite{Attia2022}. In Sec.\ \ref{sec:harmonic-repulsive-melting-line}, we determine the low-temperature melting line of the harmonic-repulsive system. In Sec.\ \ref{sec:isomorph-theory}, we develop the (zero-parameter) isomorph theory starting from a given reference state point, and in Sec.\ \ref{sec:discussions} we compare its predictions to those of  HS theories. Sec.\ \ref{sec:summary} is a summary. In Appendix \ref{appendix:temperature-dependence-of-energy} we derive the temperature dependence of the potential energy at low temperatures.

\section{The WCA and harmonic-repulsive systems}\label{sec:potentials}

\begin{figure}
	\includegraphics[width=0.99\columnwidth]{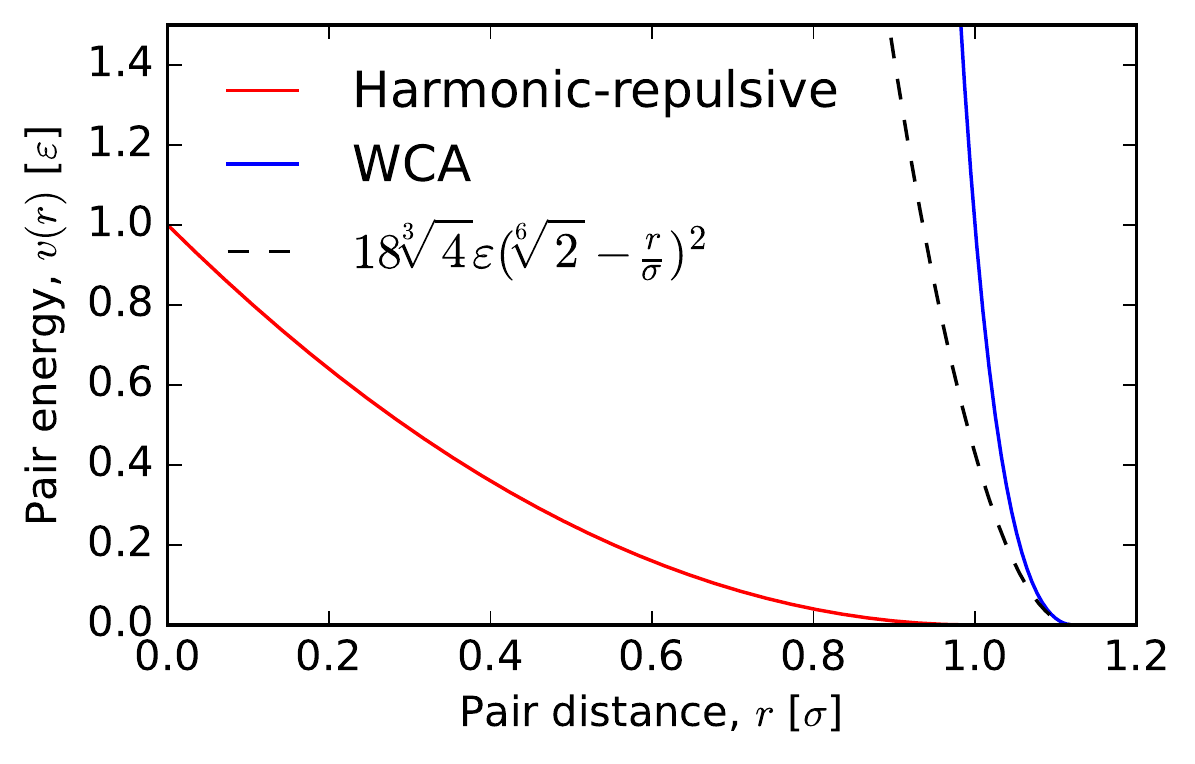}
	\caption{\label{fig:pair_potentials} The harmonic-repulsive (solid red) and \gls{WCA} (solid blue) pair potentials. The black dashed line is the harmonic-repulsive potential mapped to the same truncation distance and curvature at truncation as the \gls{WCA} pair potential.}
\end{figure}

We consider two classical mono-disperse systems of repulsive particles. Let $\bR=({\bf r}_1, {\bf r}_2, \ldots, {\bf r}_N)$ be the collective coordinate vector of $N$ particles with mass $m$ confined to the volume $V$ with periodic boundaries; the number density is given by $\rho\equiv N/V$ ($N=5120$ in all simulations). The total potential energy is a sum of pair-potential contributions,
\begin{equation}
	U({\bR}) = \sum_{i>j}^N v(|{\bf r}_j-{\bf r}_i|)\,.
\end{equation}
We shall investigate two pair-potentials. The first is the harmonic-repulsive potential defined by 

\begin{equation}\label{eq:harmonic_repulsive_pair_potential}
    v(r)=\varepsilon\left(1-\frac{r}{\sigma}\right)^2\text{ for } r<\sigma
\end{equation}
and zero otherwise. Here $\varepsilon$ has the unit of energy and $\sigma$ has the unit of length. Henceforth, quantities are reported in units derived from $m$, $\sigma$, $\varepsilon$, and the Boltzmann constant $k_B$. The red line in Fig.\ \ref{fig:pair_potentials} shows the harmonic-repulsive pair potential. This potential is often used as a model for colloidal particles, for jamming, and in coarse-grained modeling such as that behind dissipative particle dynamics \cite{Hoogerbrugge1992, Groot1997, Hern2003, Zhu2011, Espanol2017, Levashov2017, Levashov2019, Levashov2020, Santo2021}.

The second system investigated is the \gls{WCA} pair potential \cite{Weeks1971, Zhou2020} defined by
\begin{equation}\label{eq:repulsive_harmonic}
	v(r) = 4\varepsilon\left[\left(\frac{\sigma}{r}\right)^{12}-\left(\frac{\sigma}{r}\right)^{6}\right]+\varepsilon\text{ for } r<r_c
\end{equation}
and zero otherwise in which
\begin{equation}
	r_c = \sqrt[6]{2}\sigma\simeq1.1225\sigma.
\end{equation}
The \gls{WCA} potential is the standard Lennard-Jones potential cut and shifted at its minimum. This implies that the shifted-potential and shifted-force \cite{Toxvaerd2011} cutoffs are identical. The \gls{WCA} pair potential was originally introduced as the repulsive reference of the Lennard-Jones system \cite{Weeks1971, Pedersen2010}, but has since become popular on its own as a generic fluid model \cite{Heyes2006}. The blue line in Fig.\ \ref{fig:pair_potentials} shows the \gls{WCA} potential.

At low finite temperatures it is reasonable to approximate the \gls{WCA} pair potential by the first non-vanishing term of a Taylor expansion around $r=r_c$. Thus one can approximate it as follows (the dashed black line in Fig.\ \ref{fig:pair_potentials})
\begin{equation}\label{eq:v_taylor}
	v(r) \cong \frac{k_2}{2}(r_c-r)^{2}\text{ for } r<r_c
\end{equation}
and zero otherwise in which

\begin{equation}
    k_2 \equiv \left. \frac{d^2v}{dr^2} \right|_{r_c}\,,
\end{equation}
yielding $k_2=36\sqrt[3]{4}\varepsilon/\sigma^2\simeq57\varepsilon/\sigma^2$. Note that Eq.\ (\ref{eq:v_taylor}) with  $k_2=2\varepsilon/\sigma^2$ and $r_c=\sigma$ is identical to the harmonic-repulsive pair potential (Eq.\ (\ref{eq:harmonic_repulsive_pair_potential})). Thus, the physics of the two models are expected to be equivalent at low temperatures when reported in units derived from $k_2$ and $r_c$.

In the $T\to 0$ limit both pair potentials approach that of the \gls{HS} potential corrresponding to diameter $d=r_c$  \cite{Alder1957, Wood1957, Alder1959, Alder1960}, 
\begin{equation}
	v(r) = \infty \text{ for } r<d
\end{equation}
and zero otherwise. In this limit the effective HS diameter is the truncation distance, $d=r_c$, and the low-temperature solid/liquid coexistence lines approach those of the \gls{HS} system. The \gls{HS} coexistence pressure was estimated by Fernandez \textit{et al.} \cite{Fernandez2012} to
\begin{equation}\label{eq:hs_pressure}
  p_d = 11.5712(10) k_BT/d^3\,.
\end{equation}
where the value in parenthesis gives the statistical uncertainty on the last digits.
Thus when $T\to 0$ for the harmonic-repulsive system, we expect the coexistence pressure to approach (putting $d=\sigma$)
\begin{equation}\label{eq:p_bullet_hs}
  p_\bullet = 11.5712(10) k_BT/\sigma^3\,,
\end{equation}
while for the WCA system one gets with $d=\sqrt[6]{2}\sigma$
\begin{equation}
  p_\bullet = 8.1821(7) k_BT/\sigma^3\,.
\end{equation}
Throughout this paper the bullet subscript “$\bullet$” refers to the HS limit, which the potentials approach when $T\to0$, i.e., setting $d = r_c$. The fluid and solid densities were estimated by Fernandez \textit{et al.} \cite{Fernandez2012} to
\begin{equation}
    \rho^{(l)}_d = 0.938 90(5)/d^3
\end{equation}
and
\begin{equation}
    \rho^{(s)}_d = 1.037 15(9)/d^3,
\end{equation}
respectively. The densities $\rho^{(l)}_\bullet$ and $\rho^{(l)}_\bullet$ are derived by inserting the appropriate $d$'s.

The effective \gls{HS} diameter of the harmonic-repulsive and \gls{WCA} potentials are clearly smaller at finite temperatures. In this paper we develop and present analytical theories for the shape of the coexistence line when the potentials approach the HS potential.

\section{The harmonic-repulsive phase-transition lines at low temperatures}\label{sec:harmonic-repulsive-melting-line}

Recall that the solid/liquid phase transition defines a single line in the thermodynamic pressure-temperature diagram, but two lines and an in-between coexistence region in the density-temperature diagram. The \gls{WCA} and harmonic-repulsive systems are particularly simple in the sense that they, by being purely repulsive, do not have a gas-liquid phase transition, only a solid-liquid transition. The WCA phase-transition line of fluid to face-centered cubic (FCC) crystal was determined accurately in Ref.\ \cite{Attia2022} by combining the interface-pinning method \cite{Pedersen2013, Pedersen2013b, Pedersen2015} and the Gibbs-Duhem integration method \cite{Kofke1993, Kofke1993b, Frenkel2002} (the data for the coexistence line are available at \url{http://doi.org/10.5281/zenodo.6505217}). In Ref.\ \cite{Attia2022} we investigated four HS approximations for predicting the shape of the \gls{WCA} melting line. We showed that the Andersen-Weeks-Chandler HS criterion \cite{Andersen1971} gives the best predictions in the low-temperature limit. In this paper, we extend the analysis to include the predictions of the isomorph theory and also study the same problem for the harmonic-repulsive potential.

\begin{table}
\caption{\label{tab:interface_pinning} Coexistence point of the harmonic-repulsive pair potential determined by the interface-pinning method. The numbers in the parenthesis indicate the statistical uncertainty within a 95\% confidence interval.}
\begin{ruledtabular}
\begin{tabular}{ll}
Temperature, $T$ & $0.002\varepsilon/k_B$ \\
 \hline
 Coexistence pressure, $p$ & $0.02756(2)\varepsilon/\sigma^3$ \\
 Density of solid (FCC), $\rho_s$ & $1.1844(3)\sigma^{-3}$ \\
 Density of fluid, $\rho_l$ & $1.0827(2)\sigma^{-3}$ \\
 Volume difference, $\Delta v = \rho_l^{-1}-\rho_s^{-1}$ & $0.07926(4)\sigma^3$ \\
 Entropy of fusion, $\Delta s$ & $1.227(1)k_B$
\end{tabular}
\end{ruledtabular}
\end{table}

\begin{figure}
	\includegraphics[width=0.99\columnwidth]{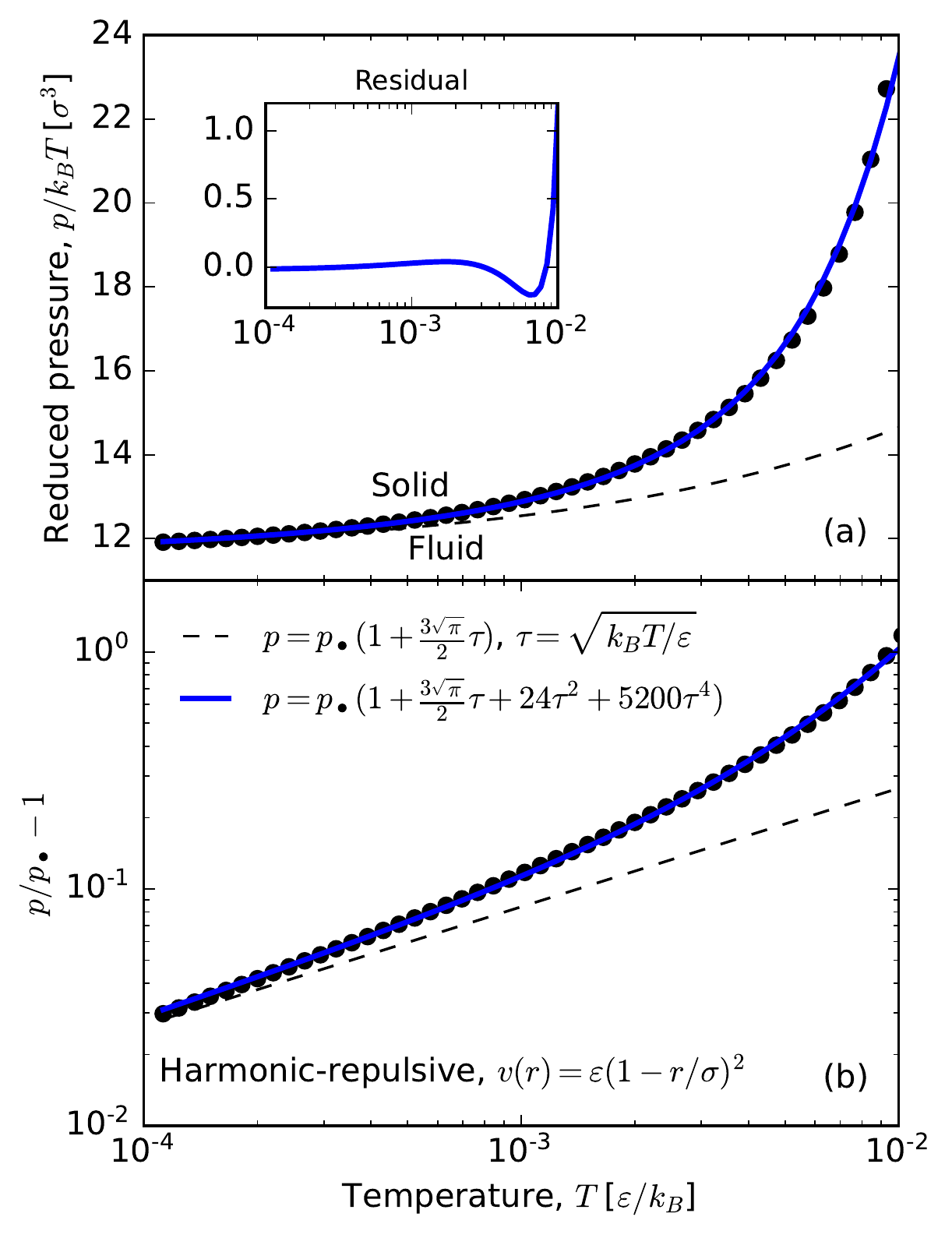}
	\caption{\label{fig:empirical_fit_pressure} Solid-fluid coexistence line of the harmonic-repulsive system in the pressure-temperature phase diagram. 
	(a) Dots show the reduced coexistence pressure, $p/k_BT$, between the fluid and FCC crystal. The black dashed line is the prediction from a HS theory (Eq.\ (\ref{eq:p_rep_harm_old})) and the blue solid line is the empirical fit Eq. (\ref{eq:empirical_fit}). The inset shows the residual of the latter. 
	(b) Dots show $p/p_\bullet-1$ on a logarithmic axis. The black dashed and blue solid lines are the same as in (a).}
\end{figure}

\begin{figure}
	\includegraphics[width=0.99\columnwidth]{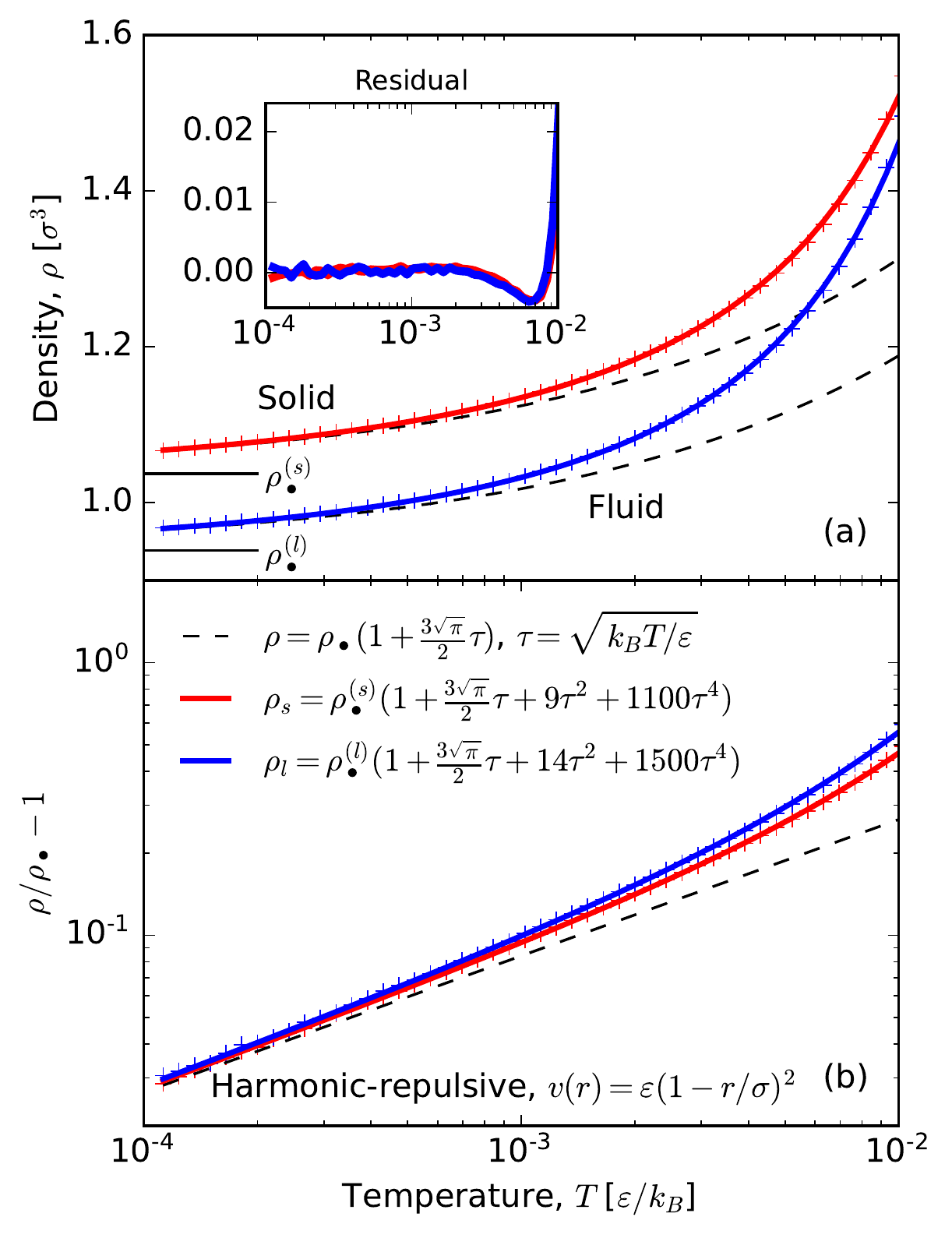}
	\caption{\label{fig:empirical_fit_density} Solid-fluid coexistence lines of the harmonic-repulsive system in the density-temperature phase diagram. 
	(a) The density of the FCC solid ($\rho=\rho_s$; red $+$'s) and the fluid ($\rho=\rho_l$; blue $+$'s) at coexistence. The black dashed lines are the low-temperature analytical predictions of the HS theory of Ref.\ \cite{Attia2022}. The blue and red solid lines are empirical fits. 
	(b) The same information as in (a) plotted as $\rho/\rho_\bullet-1$ on a logarithmic scale.}
\end{figure}

Zhu and Lu \cite{Zhu2021} computed the phase diagram of the single-component harmonic-repulsive system. This system has many crystal structures: FCC, body-centered cubic, base-centered orthorhombic, body-centered tetragonal, diamond crystal structures, and more. The phase diagram also includes several re-entrant melting regions. This richness is similar to that of other ultra-soft potentials such as the Gaussian core \cite{Stillinger1976}, EXP (exponential-repulsive) \cite{Pedersen2019}, and Hertzian-sphere pair potentials \cite{Pmies2009, Mohanty2014, Ouyang2016, Athanasopoulou2017, Muna2019, MartnMolina2019}, which are all characterized by a finite pair potential energy for $r=0$. We focus here on the low-temperature, low-density part of the phase diagram where the fluid is in equilibrium with a FCC solid. The triple-point temperature between fluid, FCC, and body-centered cubic, is given by $T_{tp}=0.012\varepsilon/k_B$ \cite{Zhu2021}, which defines the upper limit of the \emph{low-temperature} regime studied in this paper.

To accurately determine the coexistence line below $T_{tp}$ we combine the interface-pinning method \cite{Pedersen2013} at $T=0.002\varepsilon/k_B$ with numerical integration of the Clausius-Clapeyron identity \cite{Kofke1993, Kofke1993b, Frenkel2002}. The required thermodynamic inputs are estimated from simulations carried out using the leap-frog algorithm with time-step $\Delta t=0.04\sqrt{m\sigma^2/\varepsilon}$. The pressure and temperature must be constant for the interface-pinning calculations and the Clausius-Clapeyron integration. For this, we use the Langevin-dynamics algorithm of Gr{\o}nbech-Jensen and Farago \cite{GronbechJensen2014, GronbechJensen2014b} with velocity friction coefficient set to $5\varepsilon/\sigma$ and friction coefficient for the simulation-box velocity to $3.67\times10^{-7}\varepsilon/\sigma$. The simulations were conducted using the Roskilde University Molecular Dynamics (RUMD) software package version 3.6 \cite{Bailey2017} that utilizes graphics processing units (GPU) for fast computations.

Table \ref{tab:interface_pinning} gives the coexistence point determined with the interface-pinning method, and the dots in Fig.\ \ref{fig:empirical_fit_pressure}(a) show the coexistence pressure computed from the integration of the Clausius-Clapeyron identity in temperature steps given by $T_\text{next}=T_\text{previous}10^{\pm1/24}$. The dashed line is the theoretical prediction \cite{Attia2022},

\begin{equation}\label{eq:p_rep_harm_old}
  p(T) = p_\bullet\left(1+\frac{3\sqrt{\pi}}{2}\tau\right)
\end{equation}
where
\begin{equation}\label{eq:tau}
    \tau \equiv \sqrt{\frac{k_BT}{\varepsilon}}\,.
\end{equation}
In Ref.\ \cite{Attia2022} Eq.\ (\ref{eq:tau}) is derived for the \gls{WCA} potential. However, as discussed above, the physics of the WCA and harmonic-repulsive pair potentials are equivalent at low temperatures (and low pressure). Thus the theory applies equally well to the harmonic-repulsive system. The derivation of Eq. (\ref{eq:p_rep_harm_old}) is based on the Barker-Henderson \cite{Barker1967, Barker1976} and Andersen-Weeks-Chandler HS theories \cite{Andersen1971, Hansen2013}, which are identical in the low-temperature limit. Equation\ (\ref{eq:p_rep_harm_old}) is arrived at by inserting $k_2=2\varepsilon$ and $r_c=\sigma$ into Eqs. (41) and (42) of Ref.\ \cite{Attia2022}.

As a testimony to the accuracy of the theory and the computed coexistence line we note that Eq.\ (\ref{eq:p_rep_harm_old}) gives excellent agreement at low temperatures (compare the black dashed line and dots in Fig.\ \ref{fig:empirical_fit_pressure}(a)). To highlight the tiny discrepancy, Fig.\ \ref{fig:empirical_fit_pressure}(b) shows the difference to the $T\to0$ HS prediction, $p_\bullet$, by plotting $p/p_\bullet-1$ on a logarithmic axis. Even in this representation the discrepancy is barely visible at the lowest temperatures investigated.

To provide an expression valid also at high temperatures we fit the coexistence pressure to
\begin{equation}\label{eq:empirical_fit}
    p(T) = p_\bullet\left(1+\frac{3\sqrt{\pi}}{2}\tau + a_2\tau^2+a_4\tau^4\right)\,.
\end{equation}
By a least-squares fit to the data and to the above function represented as $\ln(p/p_\bullet-1)$ we find $a_2=24$ and $a_4=5200$. Equation (\ref{eq:empirical_fit}) with these parameters is shown as blue solid lines in Figs.\ \ref{fig:empirical_fit_pressure}(a) and \ref{fig:empirical_fit_pressure}(b). We note that Eq.\ (\ref{eq:empirical_fit}) is \emph{not} a theory; it is included here for reference as a practical representation of the coexistence line and not used in the remainder of the paper.

Figure \ref{fig:empirical_fit_density}(a) shows the density of the \gls{FCC} solid and fluid at coexistence. The black dashed lines show the above-mentioned HS theory applied to the coexistence density, resulting \cite{Attia2022} in
\begin{equation}
    \rho_{s}(T)=\rho^{(s)}_\bullet\left(1+\frac{3\sqrt{\pi}}{2}\tau\right)\,,
\end{equation}
and likewise for the fluid density, $\rho_{l}(T)=\rho^{(l)}_\bullet\left(1+\frac{3\sqrt{\pi}}{2}\tau\right)$. The solid lines are empirical fits similar to Eq.\ (\ref{eq:empirical_fit}), replacing pressure with density. Figure \ref{fig:empirical_fit_density}(a) shows $\rho^{(s)}/\rho^{(s)}_\bullet-1$ and $\rho^{(l)}/\rho^{(l)}_\bullet-1$ on a logarithmic scale.

\section{Isomorph-theory predictions}\label{sec:isomorph-theory}

Predictions for the thermodynamics of freezing and melting can be made within the framework of isomorph theory \cite{Pedersen2016}. Isomorphs are lines in the phase diagram along which the excess entropy $S_\text{ex}$ -- the entropy in excess of the ideal gas at the same density and temperature -- is constant. Such lines are only termed isomorphs, however, if the system obeys hidden scale invariance (Sec. I) to a good approximation because only in this case are the structure and dynamics predicted to be invariant along isomorphs. Hidden scale invariance can be checked by evaluating the virial potential-energy correlation coefficient $R$ \cite{Pedersen2008, Gnan2009} at the relevant state points: if $R$ is close to unity, the isomorph theory applies. For more information, we refer the interested reader to Refs. \onlinecite{Bailey2008, Gnan2009, Schroeder2014, Dyre2014, Dyre2018b}; it suffices here to mention that for instance the Lennard-Jones, the Yukawa (screened Coulomb), and the EXP system all have isomorphs in their liquid and solid phases. Interestingly, it was recently shown that the WCA system has isomorphs throughout its entire phase diagram because $R>\sqrt{8/3\pi}=0.92\ldots$ \cite{Attia2021}. 

In the following we apply the isomorph theory of freezing \cite{Pedersen2016} to the \gls{WCA} system.

\begin{figure}
	\includegraphics[width=0.99\columnwidth]{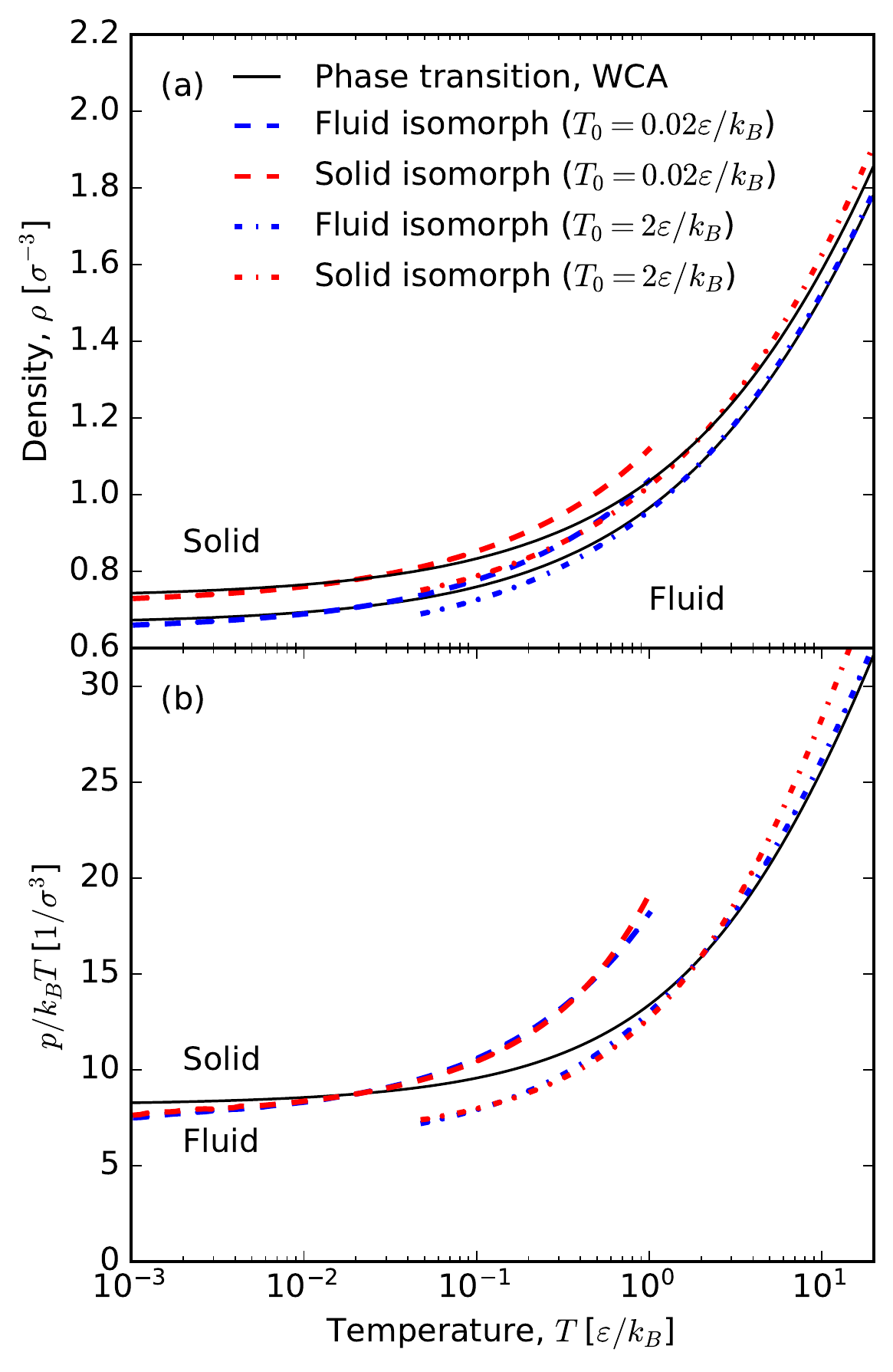}
	\caption{\label{fig:isomorph_theory_isomorphs} Isomorphs of the WCA system close to coexistence. 
        (a) shows the fluid-solid coexistence region (enclosed by solid black lines) in a $\rho$-$T$ plane, an isomorph of the fluid, and one of the solid (blue and red dashed lines, respectively). The two isomorphs cross the phase transition line at the reference temperature $T_0=0.02$. The dot-dashed lines mark the same for the reference temperature $T_0=2$.
        (b) The same as in the above panel in the ($p/k_BT$)-$T$ plane.}
\end{figure}

\begin{figure*}
	\includegraphics[width=1.99\columnwidth]{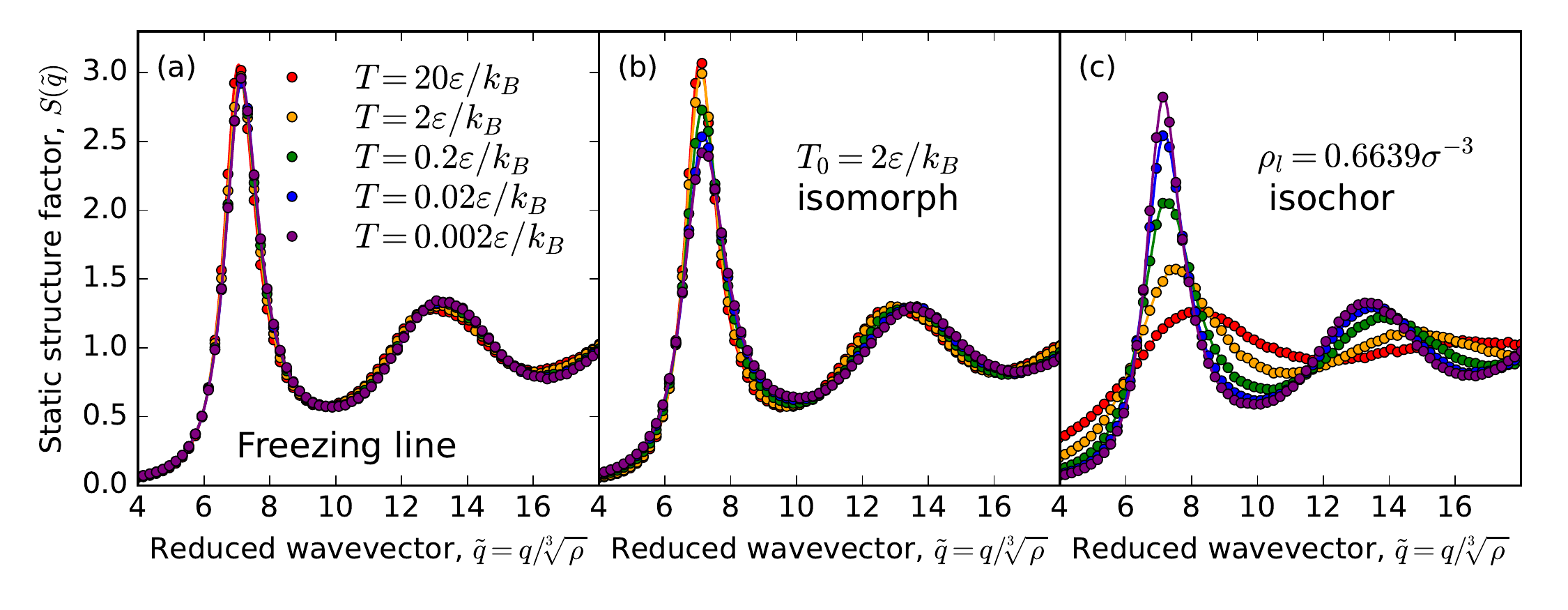}
	\caption{\label{fig:Sq} (a) The static structure factor of the \gls{WCA} system along the freezing line computed as $S({\bf q})=\langle|\sum_{n=1}^N\exp(\i {\bf q}\cdot {\bf r}_n )|^2\rangle/N$ where ${\bf q}=(0, 0, q)$. The periodic boundary condition dictates that $q=2\pi n_z/L_z$ where $n_z=1,2,\ldots$ and $L_z$ is the length of the box in the $z$-direction. $S(q)$ is shown as a function of the reduced wavevector, $\tilde q = q/\sqrt[3]{\rho}$ (this removes trivial scaling of the peak positions with density). The dots show $S(\tilde q)$ at five different state points on the freezing line. The solid lines are cubic splines serving as a guide to the eye. (b) The static structure factor along state points on the isomorph with $T_0=2\varepsilon/k$ for the same temperatures as (a). (c) The static structure factor for state points on the $\rho_l=0.6639\sigma^{-3}$ isochore for the same temperatures as in (a) and (b).}
\end{figure*}

\begin{figure}
	\includegraphics[width=0.9\columnwidth]{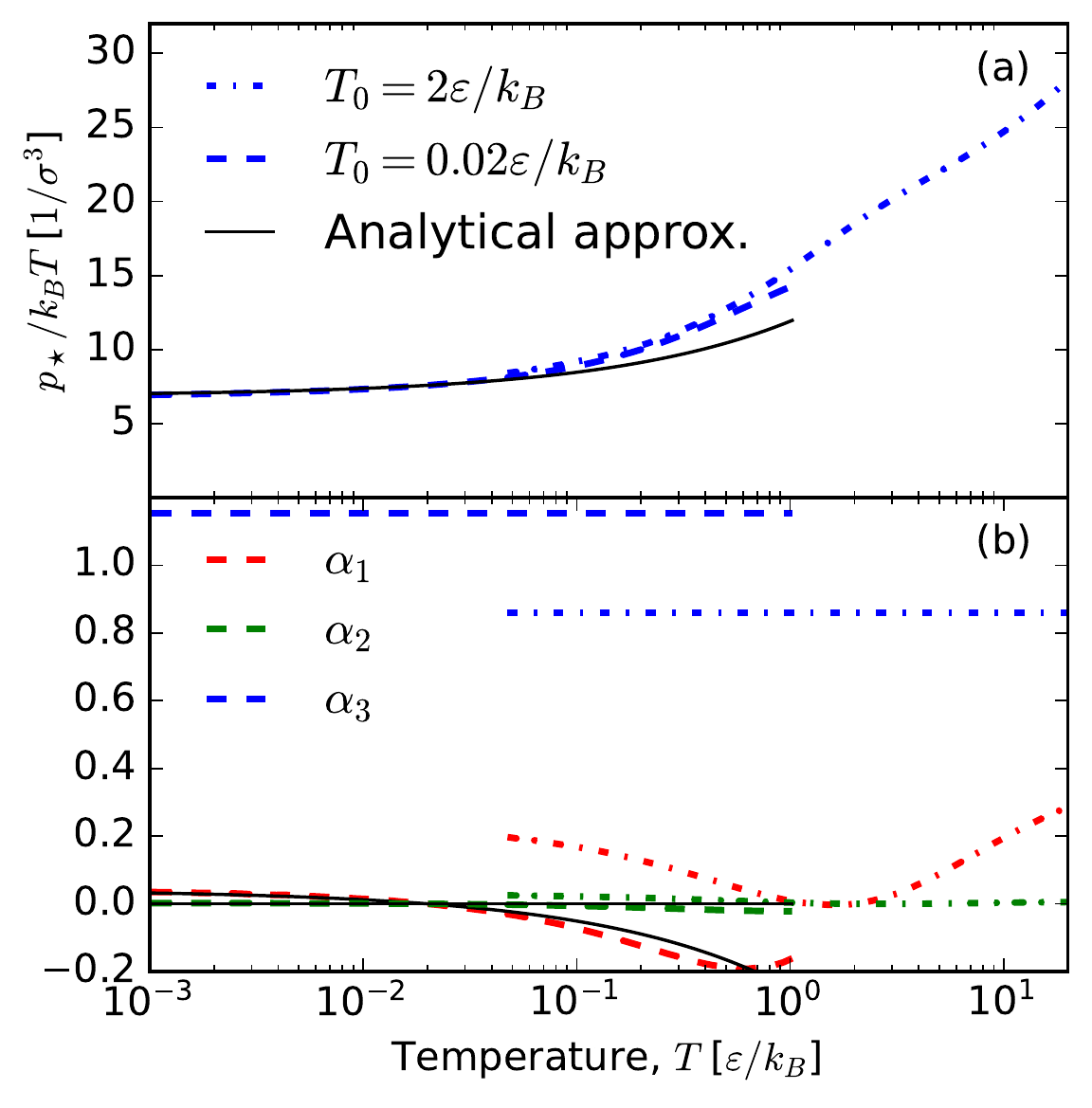}
	\caption{\label{fig:isomorph_theory_alphas}
        (a) The blue dashed line shows $p_\star$ (Eq.\ (\ref{eq:isomorph_theory_last})) evaluated using densities along the solid and fluid isomorphs with $T_0=0.02\varepsilon/k_B$ (Fig.\ \ref{fig:isomorph_theory_isomorphs}(a)). The solid black line is the analytical approximation to $p_\star$ obtained by using the densities given by Eq. (\ref{eq:density_half_harmonic}). The blue dashed-dot line is $p_\star$ evaluated using the isomorphs with $T_0=2\varepsilon/k_B$. 
        (b) The dashed lines show $\alpha$'s of Eqs.\ (\ref{eq:isomorph_theory_first})-(\ref{eq:isomorph_theory_last_alpha}) evaluated using energies and densities of the $T_0=0.02\varepsilon/k_B$ isomorphs. The solid lines are analytical approximations arrived at by insertion of Eqs.\ (\ref{eq:density_half_harmonic}) and (\ref{eq:energy_half_harmonic}) into Eqs.\ (\ref{eq:isomorph_theory_first})-(\ref{eq:isomorph_theory_last_alpha}). The dashed-dotted lines are the $\alpha$'s of the $T_0=2\varepsilon/k_B$ isomorphs. To a good approximation $|\alpha_2(T)|<|\alpha_1(T)|\ll\alpha_3$, which suggests that it is fair to assume $\alpha_1(T)=\alpha_2(T)=0$.}
\end{figure}

\begin{figure}
	\includegraphics[width=0.9\columnwidth]{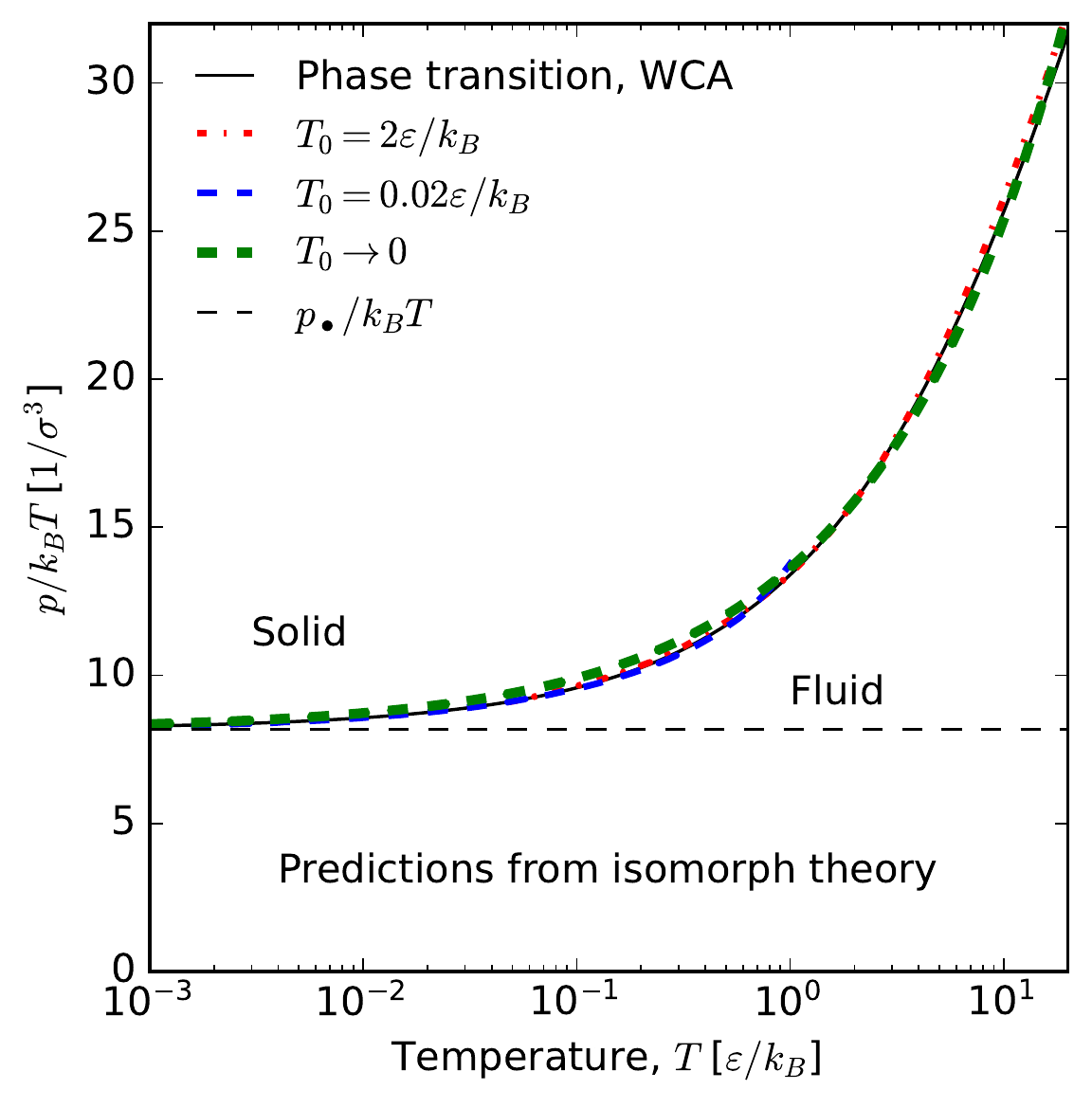}
	\caption{\label{fig:isomorph_theory} Zero-parameter predictions (colored dashed lines) for the reduced coexistence pressure (solid black), $p/k_BT$, based on the isomorph theory. The blue dashed line is Eqs. (\ref{eq:isomorph_theory})-(\ref{eq:isomorph_theory_last_alpha}) evaluated using thermodynamic data of a liquid and solid isomorph with $T_0=0.02\varepsilon/k_B$ shown as dashed lines on Fig.\ \ref{fig:isomorph_theory_isomorphs}. The red dash-dottet is constructed using isomorphs with $T_0=2\varepsilon/k_B$. The green dashed line is the analytical prediction of Eq.\ (\ref{eq:analytical_isomorph_theory}) obtained by letting $T_0\to0$ in Eqs. (\ref{eq:isomorph_theory})-(\ref{eq:isomorph_theory_last_alpha}). For comparison, the black dashed line is the prediction assuming that WCA particles are HS with the diameter at the truncation distance, $d=r_c=\sqrt[6]{2}\sigma$.}
\end{figure}

\begin{figure}
\includegraphics[width=0.9\columnwidth]{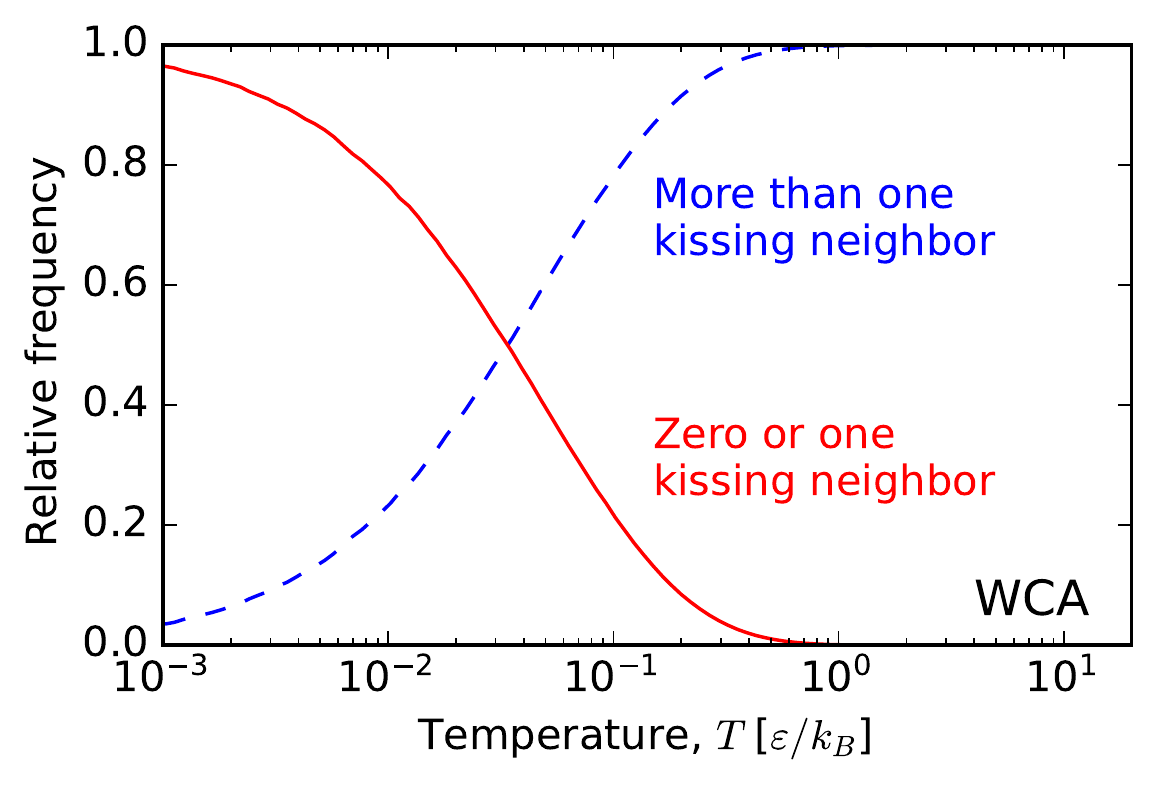}
   \caption{\label{fig:touching_neighbours} The solid red line shows the relative fraction of WCA particles with zero or one \emph{kissing} neighbors, i.e., where the pair distance is shorter than the truncation of the potential, $r_c$. The dashed blue line shows the relative frequency of particles with more than one kissing neighbor. At low temperatures, it is appropriate to assume uncorrelated particle collisions since particles have zero or one kissing neighbor.}
\end{figure}

\begin{figure}
    \includegraphics[width=0.9\columnwidth]{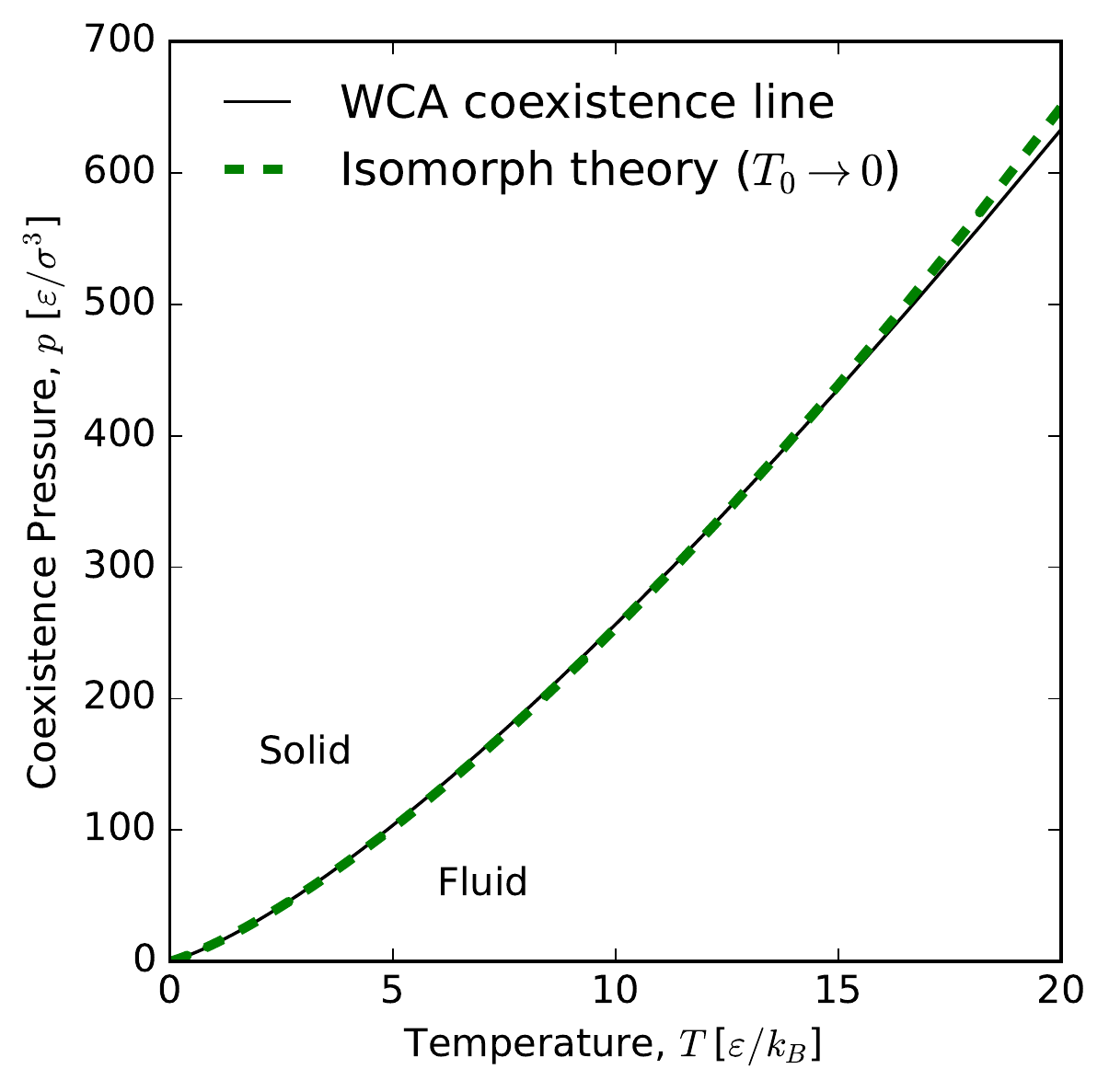}
	\caption{\label{fig:coexistence_pressure_isomorph_theory} The isomorph theory's prediction (green dashed) of the WCA coexistence pressure (solid line).}
\end{figure}

\subsection{Reference isomorphs}

An isomorph can be traced out in the thermodynamic phase diagram by numerical integration in the $\ln T$-$\ln \rho$ plane using the classic fourth-order Runge-Kutta method (RK4). At any given state point the required slope $\gamma$ defined by 
\begin{equation}
	\gamma\equiv\left(\frac{\partial\ln T}{\partial\ln \rho}\right)_{S_\text{ex}}\,,
\end{equation}
is computed from the virial ($W$) and potential-energy ($U$) fluctuations in the $NVT$ ensemble as $\gamma=\langle \Delta W\Delta U\rangle/\langle (\Delta U)^2\rangle$ \cite{Gnan2009}.

Consider a fluid and a solid isomorph both touching the coexistence line at the reference state point ($p_0$, $T_0$). Specifically, we consider below the \gls{WCA} reference state point
\begin{eqnarray}
T_0 &=& 0.02 \varepsilon/k_B,\\
p_0 &=& 0.17523\varepsilon/\sigma^3.
\end{eqnarray}
Let $u_s(T)$ and $u_l(T)$ be the average potential energy per particle and $\rho_s(T)$ and $\rho_l(T)$ the density along the solid and fluid isomorphs, respectively. The dashed lines in Fig.\ \ref{fig:isomorph_theory_isomorphs} show these two isomorphs of the WCA fluid (blue) and solid (red), respectively, crossing the coexistence line (solid black) at the reference temperature $T_0 = 0.02 \varepsilon/k_B$. The dot-dashed lines show the isomorphs with reference temperature $T_0=2\varepsilon/k_B$. Figure \ref{fig:Sq} shows the static structure factor, $S(q)$, of the fluid along the freezing line, the fluid isomorph with $T_0=2\varepsilon/k_B$, and the $\rho_l=0.6639\sigma^{-3}$ isochore. The structure is approximately invariant along the freezing line and the isomorph, but not along the isochore.

The temperature dependence of the coexistence pressure, $p(T)$, can be found by a Taylor expansion from the isomorphs to where the Gibbs free energy of the two phases are equal  \cite{Pedersen2016}. This results in
\begin{equation}\label{eq:isomorph_theory}
	p(T) = p_\star(T) [\alpha_1(T)+\alpha_2(T)+\alpha_3]
\end{equation}
in which

\begin{eqnarray}
\alpha_1(T) &=& [u_s(T)/k_BT - u_s(T_0)/k_BT_0] \label{eq:isomorph_theory_first} \nonumber \\
	   &-& [u_l(T)/k_BT - u_l(T_0)/k_BT_0], \\
\alpha_2(T) &=& \log(\rho_s(T)/\rho_s(T_0)) \nonumber \\
       &-& \log(\rho_l(T)/\rho_l(T_0)), \\
\alpha_3 &=& \frac{p_0}{k_BT_0} [\rho_l^{-1}(T_0)-\rho_s^{-1}(T_0)],\text{ and } \label{eq:isomorph_theory_last_alpha} \\
p_\star(T) &=& k_BT/[\rho_l^{-1}(T)-\rho_s^{-1}(T)]. \label{eq:isomorph_theory_last}
\end{eqnarray}
We have here redefined the $C$'s of Ref.\ \cite{Pedersen2016} and represented the same information in terms of dimensionless $\alpha$'s, introducing $p_\star$ that has the unit of pressure. The dashed lines in Fig.\ \ref{fig:isomorph_theory_alphas} show the empirical value of $p_\star$ and the $\alpha$'s along the $T_0=0.02\varepsilon/k_B$ isomorph, the green dashed line in Fig.\ \ref{fig:isomorph_theory} shows the prediction of the coexistence line by inserting these into Eq.\ (\ref{eq:isomorph_theory}). The prediction is excellent. The red dash-dot line in Fig \ref{fig:isomorph_theory} shows the prediction using the isomorphs with $T_0=2\varepsilon/k_B$. Again, the agreement is excellent.

Note that the predictions of the isomorph theory do not involve empirical fitting like Eq. (\ref{eq:empirical_fit}), merely thermodynamic information along the two isomorphs. There is, however, the freedom in picking the coexistence reference state point temperature $T_0$. In the following we remove this ambiguity by letting $T_0\to0$. This results in simple closed forms of Eqs.\ (\ref{eq:isomorph_theory})-(\ref{eq:isomorph_theory_last}), giving a good overall description of the coexistence line.

\subsection{Analytical prediction from the isomorph theory}\label{sec:analical_isomorph_theory}

To provide a closed form of Eqs.\ (\ref{eq:isomorph_theory})-(\ref{eq:isomorph_theory_last}) we need analytical expressions for the temperature dependence of the potential energy and the density along the fluid and solid isomorphs. We will do this by the mean-field approach discussed in Ref.\ \cite{Attia2021} (exact in infinite dimensions \cite{Maimbourg2016}).

First, we investigate when a mean-field is appropriate by defining a kissing neighbor as one where the pair distance is shorter than the truncation distance of the pair potential.
The solid red line in Fig.\ \ref{fig:touching_neighbours} shows the relative frequency of particles with none or a single kissing neighbor along the liquid side of the coexistence line. The analysis shows that it is reasonable to assume that particle collisions are uncorrelated at the lowest temperatures of this study since particles predominantly have none or a single kissing neighbor, i.e., collisions are uncorrelated at low temperatures.

In general, the partition function for configurational degrees of freedom is given by 
$
Z=\int_{V^N}d{\bf r}_1\ldots d{\bf r}_N\exp(-\sum_{i>j}v(r_{ij})/k_BT).
$
However, when collisions are uncorrelated, we can treat interactions in a mean-field way and write the partition function as 
\begin{equation}
    Z=Z_s^N
\end{equation}
where $Z_s$ is the partition function of a single particle moving in the potential $v_s({\bf r})$ of all other particles frozen in space. $Z_s$ has two contributions, one where the moving particle is not kissing and one where it is kissing one other particle. The former is the free volume that we will approximate by the entire volume (a low-density approximation), and the latter is $N$ times the integral
\begin{equation}\label{eq:Z_1}
Z_1 = \int_0^{r_c}4\pi r^2\exp(-v(r)/k_BT)dr.
\end{equation}
Thus, the single-particle partition function can be written as
\begin{equation}\label{eq:Z_s}
Z_s/N =Z_1+\rho^{-1}.
\end{equation}

To reach a closed form of $Z_1$ (Eq.\ (\ref{eq:Z_1})), we recall that  at low temperatures (near the phase transition), WCA particles interact only at distances close to the truncation length $r_c$, see Eq.\ (\ref{eq:v_taylor}). With this, we can get a closed form of $Z_1$ since it can be written as a Gaussian integral (see Appendix \ref{appendix:temperature-dependence-of-energy}).

We now have a mean-field theory for the partition function ($Z$) that can be used to make predictions of thermodynamic quantities. As an example, in Appendix \ref{appendix:temperature-dependence-of-energy} we derive the mean-field prediction of the energy, and in Ref.\ \cite{Attia2021} we show that
\begin{equation}
\gamma\equiv \left.\frac{d\log T}{d\log \rho}\right|_{S_\text{ex}}
\end{equation}
in the low-temperature limit is given by
\begin{equation}
	\gamma(T) = \frac{4r_c\sqrt{2k_2}}{9\sqrt{\pi k_BT}} \,\,\, (T\to 0)\,.
\end{equation}
By integration of $\gamma(T)$ using $\exp(s)\cong 1+s$ for $s\to0$ one finds
\begin{equation}\label{eq:density_half_harmonic}
\rho_l(T)=\rho_l(T_0)\left(1+\frac{9\sqrt{\pi}[\sqrt{k_BT}-\sqrt{k_BT_0}]}{2r_c\sqrt{2k_2}}\right) 
\end{equation}
and a similar expression for $\rho_s(T)$, where $\rho_s(T_0)=0.778963\sigma^{-3}$ and $\rho_l(T_0)=0.706395\sigma^{-3}$. In Appendix \ref{appendix:temperature-dependence-of-energy} we show that the low-temperature limit of the potential energy is proportional to $T^{3/2}$. Thus, using information from the reference state point only, one can write the temperature dependence of the potential energy as 
\begin{equation}\label{eq:energy_half_harmonic}
	u_l(T) = u_l(T_0)[T/T_0]^{3/2} \,\,\, (T\to 0)
\end{equation}
and a similar expression exists for $u_s(T)$ where $u_s(T_0)=5.51203\times10^{-3}\varepsilon$ and $u_l(T_0)=6.33750\times10^{-3}\varepsilon$.

\begin{figure*}
    \includegraphics[width=1.7\columnwidth]{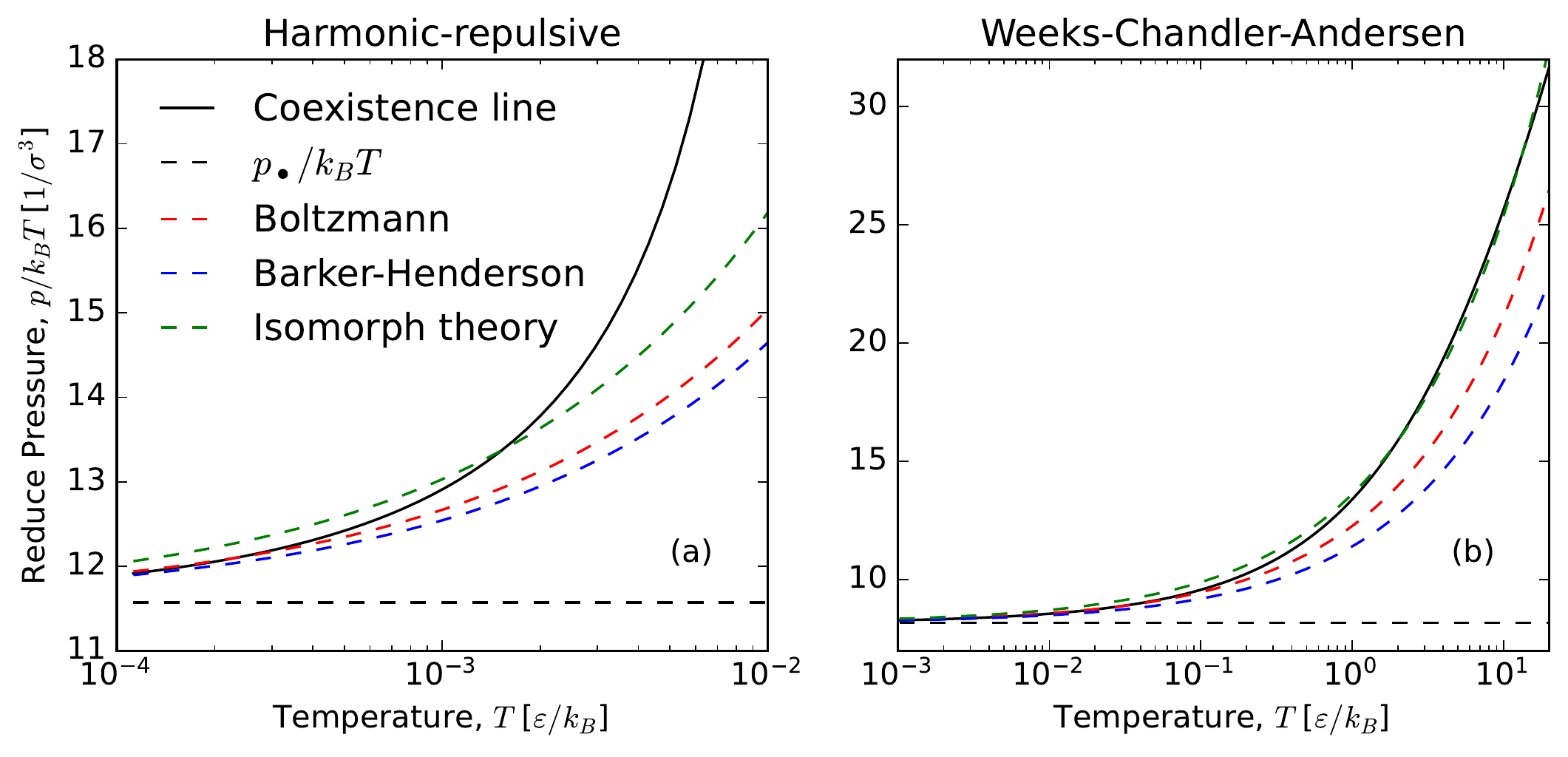}
\caption{\label{fig:coexistence_pressure}
    (a) Comparison of zero-parameter predictions (dashed lines) for the harmonic-repulsive melting line (solid curve). The black dashed line is the zero-temperature limit, $p_\bullet/k_BT$ where $p_\bullet$ is determined by assuming that the effective HS diameter is equal to the truncation distance $d=r_c=\sigma$ (Eq.\ (\ref{eq:p_bullet_hs})). The red dashed line is obtained from Boltzmann's effective HS diameter \cite{Boltzmann1890, Attia2022} (Eq.\ (\ref{eq:boltzmann})). The blue dashed line is derived from the classical HS theories \cite{Hansen2013} of Barker-Henderson \cite{Barker1967, Barker1976} and Andersen-Weeks-Chandler \cite{Andersen1971}, which are identical in the zero-temperature limit \cite{Attia2022} (Eq.\  (\ref{eq:p_rep_harm_old})). The green dashed line is the zero-parameter theory derived in Sec.\ \ref{sec:analical_isomorph_theory} (Eq.\ (\ref{eq:analytical_isomorph_theory_harmonic})). 
    (b) Comparison of the same three zero-parameter predictions for the shape of the \gls{WCA} melting line (solid black). The dashed lines are the same as in (a) applied to the \gls{WCA} system for which $r_c = \sqrt[6]{2}\sigma$ and $k_2=36\sqrt[3]{4}\varepsilon/\sigma^3$.}
\end{figure*}

The solid black lines in Fig.\ \ref{fig:isomorph_theory_alphas} show the analytical approximations of $p_\star$, $\alpha_1$, and $\alpha_2$ ($\alpha_3$ is a constant that does not need to be evaluated). We note that $|\alpha_2(T)|<|\alpha_1(T)|\ll\alpha_3$. 
Putting $\alpha_1(T)=\alpha_2(T)=0$ Eqs. (\ref{eq:isomorph_theory})-(\ref{eq:isomorph_theory_last}) reduce to

\begin{equation}\label{eq:semiempirical}
p(T) = \frac{p_0 T}{T_0}\left(1+\frac{9\sqrt{\pi}[\sqrt{k_BT}-\sqrt{k_BT_0}]}{2r_c\sqrt{2k_2}}\right)
\end{equation}
after insertion of Eq.\ (\ref{eq:density_half_harmonic}).
The low-temperature HS limit can be used as a reference point, i.e., $T_0\to0$. Equation\ (\ref{eq:semiempirical}) then simplifies to

\begin{equation}\label{eq:analytical_isomorph_theory}
	p(T) = p_\bullet\left(1+\frac{9\sqrt{\pi k_BT}}{2r_c\sqrt{2k_2}}\right)\,.
\end{equation}
This prediction is shown as green dashed lines in Figs. \ref{fig:isomorph_theory} and \ref{fig:coexistence_pressure_isomorph_theory}. It compares well to the \gls{WCA} coexistence line that is shown in solid black. In the following we will argue, though, that while the agreement is expected at low temperatures, the good overall agreement may result from a cancellation of errors of the assumptions at higher temperatures. 

For the harmonic-repulsive pair potential, the prediction of the isomorph theory is

\begin{equation}\label{eq:analytical_isomorph_theory_harmonic}
    p(T)=p_\bullet\left(1+\frac{9\sqrt{\pi}}{4}\tau\right)\,,
\end{equation}
which is derived by insertion of $k_2=2\varepsilon/\sigma^2$ and $r_c=\sigma$ into Eq. (\ref{eq:analytical_isomorph_theory}), where $\tau=\sqrt{k_BT/\varepsilon}$ (Eq. (\ref{eq:tau})).

\begin{figure}
    \includegraphics[width=0.85\columnwidth]{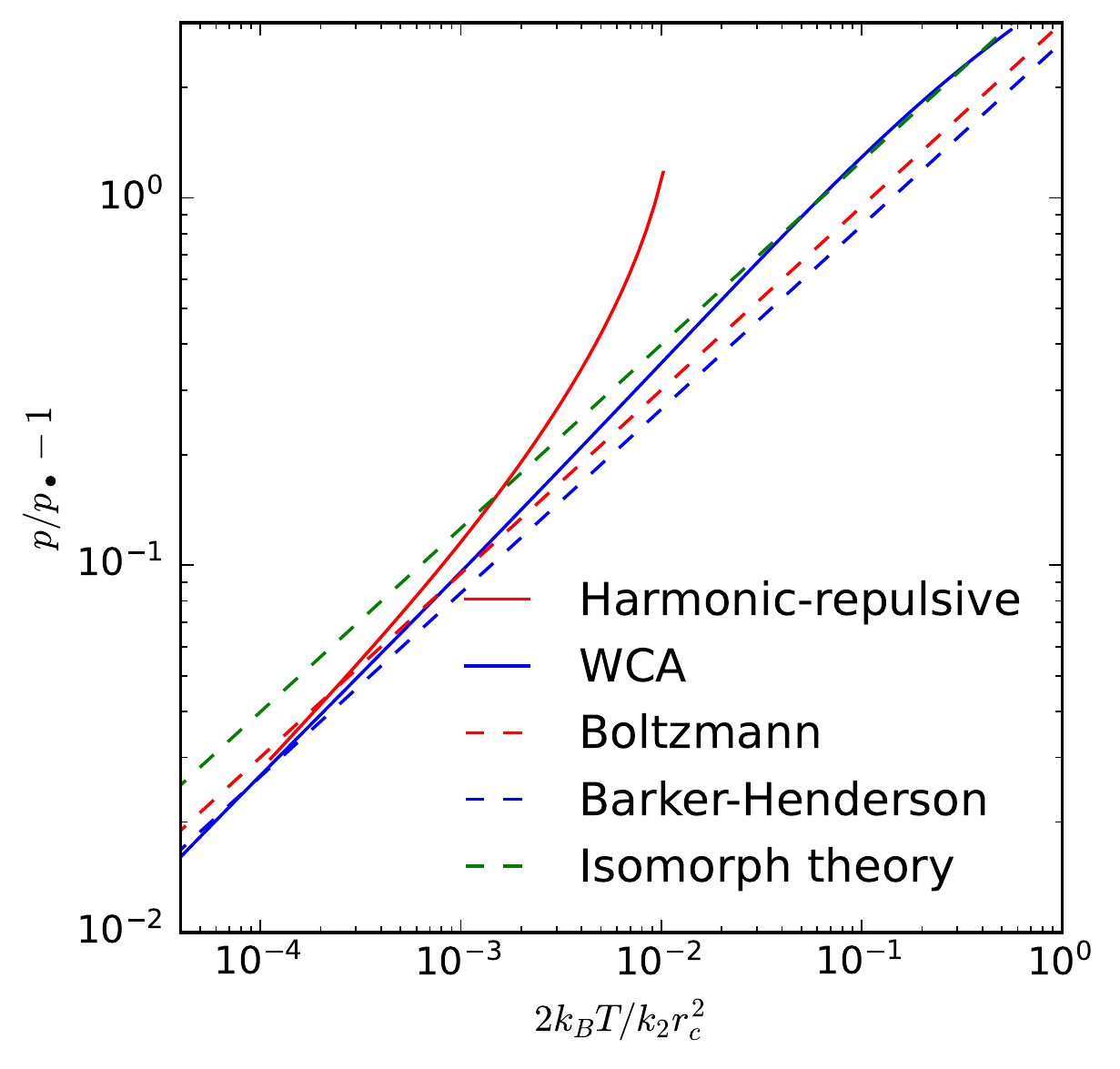}
	\caption{\label{fig:coexistence_pressure_scaled} Comparison of zero-parameter theories (dashed lines) for the harmonic-repulsive (solid red) and \gls{WCA} (solid blue) melting lines. This figure contains the same information as Fig.\ \ref{fig:coexistence_pressure} represented as $p/p_\bullet-1$ versus $2k_BT/k_2r_c^2$.}
\end{figure}

\section{Discussion}\label{sec:discussions}
The \gls{WCA} melting line based on isomorph theory gives an excellent fit to data over four orders of magnitude in temperature (Fig.\ \ref{fig:isomorph_theory}). Two essential assumptions enter into the deviation:

\begin{description}
    \item[(A)] Particle collisions are uncorrelated (Fig.\ \ref{fig:touching_neighbours});
    \item[(B)] The \gls{WCA} potential can be approximated by the harmonic-repulsive potential (Eq. (\ref{eq:v_taylor}); Fig.\ \ref{fig:pair_potentials}).
\end{description}%
Figure \ref{fig:touching_neighbours} show that the former is valid at $T<0.03k_B/\varepsilon$ (pragmatically defined as where the red and blue lines cross in Fig.\ \ref{fig:touching_neighbours}).

To analyze the validity of the latter approximation we conduct a comparative study of the two modes. To this aim, Fig.\ \ref{fig:coexistence_pressure}(a) shows the isomorph theory prediction (green dashed) of the harmonic-repulsive melting line (solid black). In the same figure, the red dashed line is the zero-parameter prediction obtained by inserting into Eq.\ (\ref{eq:hs_pressure}) Boltzmann's effective HS criterion 
\cite{Boltzmann1890, Attia2022},
\begin{equation}
    v(d) = k_BT
\end{equation}
leading to
\begin{equation}
    d = \sigma(1-\sqrt{k_BT/\varepsilon}),
\end{equation}
In the low-temperature, the prediction for the coexistence pressure is
\begin{equation}\label{eq:boltzmann}
    p(T)=p_\bullet(1+3\tau)\text{ for } T\to 0\,.
\end{equation}
The blue dashed line is the low-temperature HS prediction derived from Barker-Henderson/Andersen-Weeks-Chandler assumptions (Eq.\ (\ref{eq:p_rep_harm_old})). Clearly, all theories underestimate the value of the coexistence pressure at the highest temperatures of this study. This is attributed to errors related to assumption (A), i.e., that particle collisions are uncorrelated.

Figure \ref{fig:coexistence_pressure}(b) shows the three zero-parameter theories for the \gls{WCA} melting line. To compare the results of the three theories and two models, Fig.\ \ref{fig:coexistence_pressure_scaled} collects all the information of Fig.\ \ref{fig:coexistence_pressure} in a double-logarithmic plot with $2k_BT/k_2r_c^2$ along the abscissa and $p/p_\bullet-1$ along the ordinate. In this representation, the zero-parameter theories are straight lines with slope 1/2 and the coexistence lines of the two pair-potential collapse at low temperatures. However, the temperatures for which the two models give the same melting line (in reduced units) are rather low, and for higher temperatures, the WCA potential has a lower coexistence pressure since it is harder than the repulsive harmonic potential (compare the solid blue to and the black dashed lines in Fig.\ \ref{fig:pair_potentials}).
Moreover, compared to the harmonic-repulsive melting line, the theories predict a lower pressure. As mentioned, this low pressure likely results from the ignored many-body effects, i.e., from the above mention assumption (A). This suggests that the good overall isomorph predictions result in part from the cancellation of assumptions (A) and (B).

\section{Summary}\label{sec:summary}

We have developed an isomorph-theory-based zero-parameter prediction for the melting lines of the harmonic-repulsive and \gls{WCA} potentials. The new theory generalizes the idea of an effective one-dimensional phase diagram from HS theories without referring to a specific reference potential. In effect, the shape of the coexistence line can be evaluated using any of its points as reference state point. We have shown how to generate an analytical prediction using the theory for the shape of the isomorph presented in Ref.\ \cite{Attia2021}. Alternatively, we note that Bøhling \textit{et al.} \cite{Bohling2014} provided an approach to get an analytic expression for the shape of the isomorph for any pair-potential, which provides an alternative way to arrive at a closed-form expression of the melting line. We leave this line of reasoning for the future.

\section*{Acknowledgment}
This work was supported by the VILLUM Foundation’s
Matter grant (No. 16515).

\section*{Data availability}
The data that support the findings of this study are openly available in Zenodo at \url{http://doi.org/10.5281/zenodo.7646884}, reference number 7646884.

\appendix

\section{Temperature dependence of energy}\label{appendix:temperature-dependence-of-energy}
In the main part of the paper, we use the approximation $u(T)\propto T^\frac{3}{2}$ at low temperatures. Here, that approximation is justified.

In Ref.\ \onlinecite{Attia2021} we show that in the low-density, the low-temperature limit of a pairwise quantity $A(r)$, which is zero for $r>r_c$, has an expectation value that is computed as is 
\begin{equation}\label{eq:general-expectation-value}
    \langle A\rangle = \frac{1}{Z_s(\rho, T)} \int_0^{r_c} A(r)p(r)dr
\end{equation}
where 
\begin{equation}
    p(r)=4\pi r^2\exp[-v(r)/k_BT]
\end{equation}
is the unnormalized probability,
\begin{equation}
Z_s(\rho, T)/N=Z_1(T)+\frac{1}{\rho}
\end{equation}
(see Eq.\ (\ref{eq:Z_s})) and
\begin{equation}
    Z_1=\int_0^{r_c}p(r)dr.
\end{equation}
To give an approximation of 
\begin{equation}
 u(T)=\langle v\rangle   
\end{equation}
for the pair potential $v(r)=\frac{k_2}{2}(r_c-r)^2$ we need to evaluate integrals of the form
\begin{eqnarray}   
    I_m(T) &=& \int_0^{r_c} [v(r)]^m p(r)dr\\
    &=& \frac{4\pi k_2^m}{2^m}\int_0^{r_c} (r_c-r)^{2m} r^2\exp\left[-\frac{k_2(r_c-r)^2}{2k_BT}\right]dr \nonumber
\end{eqnarray}
From Eq.\ (\ref{eq:general-expectation-value}) the expectation value of the energy per particle ($\langle v\rangle$=$\langle U\rangle/N$) can be written as
\begin{equation}\label{eq:v2-fraction}
    \langle v\rangle = \frac{I_1(T)}{I_0(T)+\frac{1}{\rho}}\,.
\end{equation}
By substitution of 
\begin{equation}
    t=(r_c-r)^2/\tilde T
\end{equation}
where 
\begin{equation}
\tilde T\equiv2k_BT/k_2
\end{equation}
so $r=r_c-\tilde T^{\frac{1}{2}}t^{\frac{1}{2}}$ and $dr=-\frac{1}{2}\tilde T^{\frac{1}{2}}t^{-\frac{1}{2}}dt$, we can write the integral as
\begin{widetext}
\begin{equation}
    I_m(T) = \frac{4\pi k_2^m}{2^{m+1}}\int_0^{r_c^2/\tilde T}\left[ 
    r_c^2 \tilde T^{m+\frac{1}{2}}t^{m-\frac{1}{2}}
    +\tilde T^{m+\frac{3}{2}}t^{m+\frac{1}{2}}
    +2r_c\tilde T^{m+1}t^{m}
    \right]\exp(-t)dt
\end{equation}
\end{widetext}
In the low-temperature limit, $\tilde T\to0$, the integral approaches
\begin{equation}
    I_m(T) = \frac{4\pi r_c^2k_2^m}{2^{m+1}}\tilde T^{m+\frac{1}{2}}\int_0^\infty t^{m-\frac{1}{2}}\exp(-t)dt\,.
\end{equation}
Thus, the integral of the above approaches the gamma function,
\begin{equation}
\Gamma(z)\equiv\int_0^\infty t^{z-1}\exp(-t)dt\,,
\end{equation}
leading to
\begin{equation}
        I_m(T) = \frac{4\pi r_c^2k_2^m}{2^{m+1}}\tilde T^{m+\frac{1}{2}}\Gamma\left(m+\frac{1}{2}\right)\,.
\end{equation}
From this we get 
\begin{equation}
I_0(T) = 2\pi r_c^2\sqrt{\frac{2\pi k_BT}{k_2}}
\end{equation}
and
\begin{equation}
I_1(T) = \frac{r_c^2k_2}{2}\left(\frac{2\pi k_BT}{k_2}\right)^\frac{3}{2}.
\end{equation}
At low temperatures, the denominator of Eq.\ (\ref{eq:v2-fraction}) approaches $1/\rho$ since $I_0(T)\propto\sqrt{T}$. The change of $\rho$ in this study is in the order of 10\% while the enumerator of Eq.\ (\ref{eq:v2-fraction}), $I_1(T)\propto T^\frac{3}{2}$ changes orders of magnitude. This justifies the approximation
\begin{equation}
    u(T) \propto T^\frac{3}{2}
\end{equation}
used in the main part of the paper.

\bibliography{references}

\end{document}